\documentclass[aps,prl,a4,reprint,showpacs,twoside,floatfix,superscriptaddress]{revtex4-1}

\usepackage{multirow}
\usepackage{graphicx}
\usepackage{dcolumn}
\usepackage{bm}
\usepackage{times}
\usepackage{mathptmx}
\usepackage{color}
\usepackage{amsmath}
\usepackage{amsfonts}
\usepackage[T1]{fontenc}
\usepackage{color}

\begin{document}

\title{Ultrafast Modification of the Polarity at LaAlO$_3$/SrTiO$_3$ Interfaces}

\author{A. Rubano}
 \affiliation{Dipartimento di Fisica, Universit\`{a} di Napoli ``Federico II'', Compl.\ Univ.\ di Monte S.\ Angelo, v.\ Cintia, 80126 Napoli, Italy}

\author{T. G\"{u}nter}
 \affiliation{Helmholtz-Institut f\"{u}r Strahlen- und Kernphysik, Universit\"{a}t Bonn, Nussallee 14 - 16, 53115 Bonn, Germany}

\author{M. Fiebig}
 \affiliation{Helmholtz-Institut f\"{u}r Strahlen- und Kernphysik, Universit\"{a}t Bonn, Nussallee 14 - 16, 53115 Bonn, Germany}
 \affiliation{Department of Materials, ETH Z\"{u}rich, Vladimir-Prelog-Weg 4, 8093 Zurich, Switzerland}

\author{F. Miletto Granozio}
 \affiliation{Dipartimento di Fisica, Universit\`{a} di Napoli ``Federico II'', Compl.\ Univ.\ di Monte S.\ Angelo, v.\ Cintia, 80126 Napoli, Italy}
\affiliation{CNR-SPIN, Compl.\ Univ.\ di Monte S.\ Angelo, v.\ Cintia, 80126 Napoli, Italy}

\author{L.~Marrucci}
 \affiliation{Dipartimento di Fisica, Universit\`{a} di Napoli ``Federico II'', Compl.\ Univ.\ di Monte S.\ Angelo, v.\ Cintia, 80126 Napoli, Italy}
\affiliation{Institute of Applied Science and Intelligent Systems-ISASI, Via Campi Flegrei 34, 80078 Pozzuoli}

\author{D. Paparo}
 \thanks{d.paparo@isasi.cnr.it}
 \affiliation{Institute of Applied Science and Intelligent Systems-ISASI, Via Campi Flegrei 34, 80078 Pozzuoli}

\maketitle
\label{sec:main}

\textbf{Oxide growth with semiconductor-like accuracy has led to atomically precise thin films and interfaces that exhibit a plethora of phases and functionalities not found in the oxide bulk material. This yielded spectacular discoveries such as the conducting, magnetic or even superconducting LaAlO$_3$/SrTiO$_3$ interfaces separating two prototypical insulating perovskite materials. All these investigations, however, consider the static state at the interface, although studies on fast oxide interface dynamics would introduce a powerful degree of freedom to understanding the nature of the LaAlO$_3$/SrTiO$_3$ interface state. Here we show that the polarization state at the LaAlO$_3$/SrTiO$_3$ interface can be optically enhanced or attenuated within picoseconds. Our observations are explained by a model based on charge propagation effects in the interfacial vicinity and transient polarization buildup at the interface.}
\\

Interfaces between transition metal oxides exhibit unique functionalities~\cite{Hwang}. One of the most fascinating examples was the discovery of a high-mobility two-dimensional electron gas (2DEG) at the interface between two insulating perovskites, LaAlO$_3$ and SrTiO$_3$ (LAO/STO)~\cite{Ohtomo}. The 2DEG formation is a threshold process: For $n\geq4$ monolayers of LAO on STO the interface becomes metallic~\cite{Thiel} and even superconducting~\cite{Li}. At $n=3$ an electric field~\cite{Thiel} or light~\cite{Huijben2006} control the phase at the LAO/STO interface so that a \textit{reversible} insulator-to-metal transition (IMT) can occur. This may serve as basis of optoelectronics applications~\cite{Irvin2010, Tebano2012}, yet with optical control on the timescale of seconds or longer.

In contrast, studies on fast oxide interface dynamics are still rare although such investigations would be invaluable to understanding the nature of the LAO/STO interface state: Time dependence allows one to separate and recognize the various interactions contributing to the interfacial ground state via the different inherent timescales on which they respond after optical excitation. For example, transient ultrafast absorption spectroscopy led to evidence for a strong influence of the LAO film on self-trapped polaron formation at the LAO/STO interface~\cite{Yamada2013}.

Here, we reveal the mechanisms guiding build-up and depletion of photoinduced polarity at the LAO/STO interface. We obtain an optically induced sub-picosecond \textit{enhancement or attenuation} of interface polarity; both may exceed $50$\%. This effect is established by three competing photocarrier mechanisms: screening, asymmetric drift and trapping. We track the LAO/STO carrier dynamics by all-optical time-resolved pump-probe spectroscopy, using second harmonic generation (SHG) as probe process which detects the interfacial inhomogeneity while masking the bulk.

LAO films of different thicknesses were deposited by pulsed-laser deposition on TiO$_2$-terminated STO(001) substrates with unit-cell control of the film thickness by high-energy electron diffraction, as described in Ref.~\onlinecite{Rubano2011} and Supplemental Material (SM). All samples with a LAO thickness above the threshold of 4 unit cells show the same qualitative dynamics. Therefore, in this Letter, we focus our attention on two representative samples with a LAO film of two (LS2) and six (LS6) unit cells. LS2 and LS6 represent the LAO/STO samples with initial insulating and conducting interfaces, respectively. Our pump beam excites photocarriers at 4.35~eV, i.e., above the direct band gap of STO of about 3.75~eV~\cite{Goldschmidt1987}, but not across the bulk band gap of LAO, so that most of the observed dynamics resides in STO. We therefore include the air/STO interface dynamics of a TiO$_2$-terminated STO(001) substrate in our investigation.

Optical SHG, i.e., doubling the frequency of a light wave, has emerged as a powerful tool for investigating oxide thin films, heterostructures and interfaces~\cite{Tokura2007,Savoia2009,Ogawa2009,Rubano2011,Rubano2013,Rubano2013a,Trassin2015}. The SHG intensity is given by $I_{\rm SHG}\propto |\mathbf{E}(2\omega)|^2$ with $\mathbf{E}(2\omega)\propto\overleftrightarrow{\chi}(\omega,\omega,2\omega):~\mathbf{E}(\omega)\mathbf{E}(\omega)$ and $\mathbf{E}(2\omega)$ and $\mathbf{E}(\omega)$ as electric fields of the light waves at $2\omega$ and $\omega$, respectively. $\overleftrightarrow{\chi}(\omega,\omega,2\omega)$ is the second order susceptibility tensor. It is non-zero when inversion symmetry is broken. SHG is therefore ideal for probing interfaces and surfaces, where this violation occurs inherently. In the LAO/STO and STO/air systems, the SHG amplitude scales with the interfacial polarity while the centrosymmetric bulk constituents do not contribute to the SHG signal~\cite{Shen1994}.

Previous static SHG experiments provided valuable insight into the LAO/STO system. We probed the behavior of the STO at the interface with SHG-photon energies up to 4 eV. After verifying that there are no contributions from the STO bulk, the LAO, or the LAO/air interfaces~\cite{Savoia2009,Rubano2011}, we identified two components of the SHG susceptibility, here denominated as $\chi_{\rm loc}$ and $\chi_{\rm ext}$, that completely describe the SHG spectroscopic features observed at the LAO/STO and STO/air interfaces~\cite{Rubano2011,Rubano2013a}. The term $\chi_{\rm loc}$ couples to the transition from the O$^{2-}(2p)$-dominated STO valence band to the lowest STO conduction sub-bands, mainly formed by the Ti$^{4+}(3d_{xy})$ orbital. In complement, $\chi_{\rm ext}$ represents the transition to the Ti$^{4+}(3d_{xz,yz})$ orbitals. Since the subbands with a $d_{xy}$ character are more localised at the interface than those with a $d_{xz,yz}$ character~\cite{Popovic2008,Delugas2011,Plumb2014}, $\chi_{\rm loc}$ predominantly probe the polarity of the very first interfacial atomic layers whereas $\chi_{\rm ext}$ probes a deeper region, typically of several atomic layers. Comparison of SHG from $\chi_{\rm loc}$ and $\chi_{\rm ext}$ allowed us to disentangle the particular role of the interfacial Ti$^{4+}(3d)$ orbitals~\cite{Rubano2011,Rubano2013a,Paparo2013}.

\begin{figure}[t]
 \includegraphics[angle=0,scale=0.46]{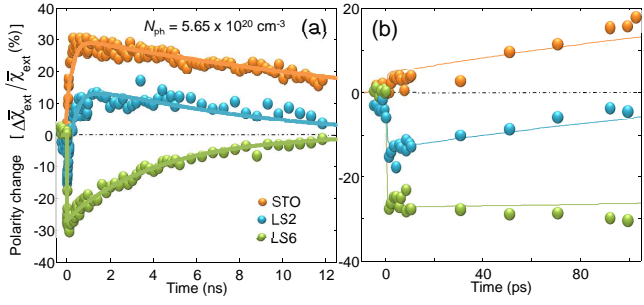}
 \caption{(a, b) Temporal evolution of $\Delta \bar{\chi}_{\rm ext}/\bar{\chi}_{\rm ext}$ for $N_{\text{ph}}=5.65\cdot10^{20}$ cm$^{-3}$. Solid lines are fits (see text for details). Note the striking difference of the dynamics between insulating and conductive samples that points to a complex interplay of different photoinduced mechanisms as better explained in the text.}\label{fig1}
\end{figure}

This insight renders SHG the ideal probe for the present time-resolved pump-probe study on the carrier dynamics around the LAO/STO interface. For this purpose, we integrate the SHG susceptibilities along $z$, which yields $\bar{\chi}_{\rm loc/ext}$, and we consider their temporal variation $\Delta\bar{\chi}_{\rm loc/ext}$ with $\Delta\bar{\chi}_{\rm loc/ext}=0$ for $t < 0$. As better explained in SM, in our derivation we neglect any effect due to the transient redistribution of the electronic populations by the pump beam, also known as state-filling effect~\cite{Sabbah2002}, since the photoinduced reflectivity changes are approximately the same for conductive and insulating samples, as reported in Fig. 2 of SM. Similarly, we assume that the Fresnel factors vary too little for explaining the absolute SHG signal variations and, in any case, they vary approximately in the same way for all the samples in order to explain the different dynamics at a given laser fluence.

Figure~\ref{fig1}(a) and its zoomed view in Fig.~\ref{fig1}(b) show the time-dependant change of $\chi_{\rm ext}$ in the LS2 and LS6 samples and STO at a pump laser intensity of 1~mJ cm$^{-2}$ (excitation density $N_{\rm ph}=5.65\cdot 10^{20}$~cm$^{-3}$). The samples show an initial sub-ps drop of the SHG signal, followed by a recovery on the 100-ps timescale and restoration of the initial state over nanoseconds. With increasing LAO coverage, $\Delta\bar{\chi}_{\rm ext}/\bar{\chi}_{\rm ext}$ evolves from positive to negative values. In LS2 and STO the ratio $\Delta\bar{\chi}_{\rm ext}/\bar{\chi}_{\rm ext}$ becomes positive, implying a transient \textit{enhancement} of the polarity of these samples by the photoexcitation whereas in LS6 the relation $\Delta\bar{\chi}_{\rm ext}/\bar{\chi}_{\rm ext}<0$ indicates a transient \textit{attenuation} of the polarity. Strikingly, the SHG susceptibility changes may amount to as much as $\pm 30$\%.

\begin{figure}[t]
 \includegraphics[angle=0,scale=0.48]{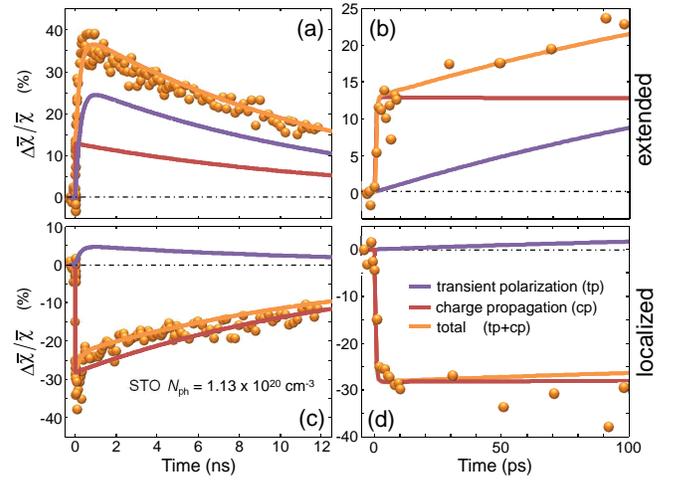}
 \caption{Time dependence of the relative change of SHG signal in STO after an optical excitation generating a photocarrier density of $N_{\rm ph}=1.13\cdot10^{20}$ cm$^{-3}$ (data points). The fits reproduce this measured change (yellow line) and its respective contributions from the \textit{charge-propagation} (`cp', red line) and \textit{transient-polarization} (`tp', blue line) mechanisms. SHG susceptibilities $\bar{\chi}$ probe the local state at the interface (panels c,d) and the extended environment around it (panels a,b), respectively.}\label{fig2}
\end{figure}

\begin{figure}
 \includegraphics[angle=0,scale=0.33]{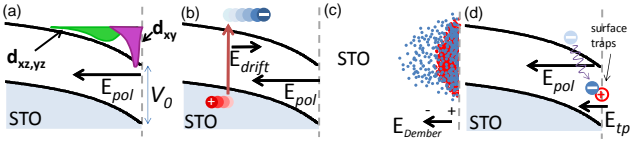}
 \caption{Nonequilibrium charge dynamics near the LAO/STO interface. (a) Band-bending and different extension of the charge concentration associated, respectively, to the $d_{xy}$-like and $d_{xz,yz}$-like subbands of STO at the interface (as adapted from Refs.~\onlinecite{Delugas2011,Deluca2014}). We notice that the medium beyond the dashed line may be either air, like in bare STO, or a LAO over-layer. (b) Screening drift as first process contributing to the \textit{charge-propagation} mechanism. (c) Photo-Dember effect as second process contributing to the \textit{charge-propagation} mechanism (as adapted from Ref.~\onlinecite{wikiDember}). (d) Interfacial charge trapping as process contributing to the \textit{transient-polarization} mechanism.}
\label{fig5}
\end{figure}

\begin{figure}[t]
 \includegraphics[angle=0,scale=0.45]{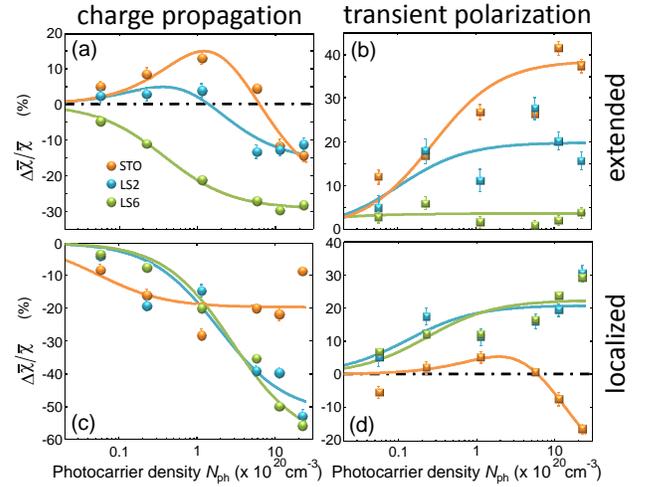}
 \caption{Dependence of the relative change of SHG susceptibility of the \textit{charge-propagation} and \textit{transient-polarization} contributions as defined in Eq.~(\ref{eq3}) on the density of the optical excitation $N_{\rm ph}$. Symbols represent fits of Eq.~(\ref{eq3}) to measured data as in Fig.~\ref{fig2}. The amplitudes have been normalized to the corresponding value of $\bar{\chi}$ at time $t<0$. Solid lines are fits (see text). Here and in Fig.~\ref{fig3} error bars indicate statistical errors at 68.3\% of confidence level obtained through a chi-square test as described in Ref.~\onlinecite{Numerical}. The error bars are missing when their extension is not larger than symbol size.}
\label{fig4}
\end{figure}

\begin{figure}
 \includegraphics[angle=0,scale=0.66]{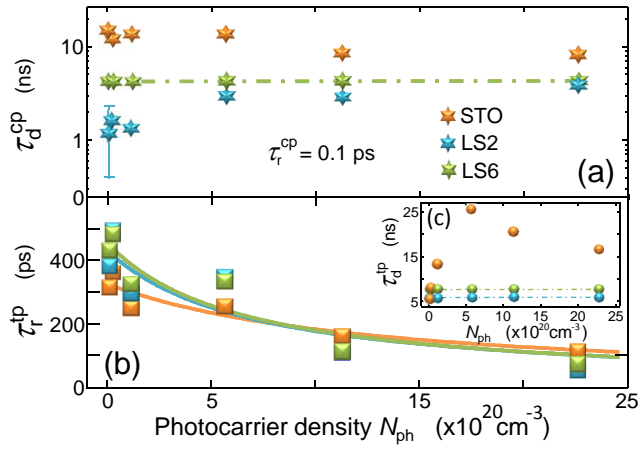}
 \caption{Dependence of the relaxation constants defined in Eq.~(\ref{eq3}) on the density of the optical excitation in the three types of samples. Solid lines are fits (see text). Dot-dashed lines indicate values of relaxation constants that did not reveal a dependance on the photocarrier density and were therefore replaced with a single fitting parameter over all the photocarrier density range.}
\label{fig3}
\end{figure}

Before proceeding with the data analysis, we need to identify the different types of contributions constituting the dynamics expressed by Fig.~\ref{fig1}. We will focus this discussion on the simplest system, bare STO as depicted in Fig.~\ref{fig2}, from which we will move on to further investigation of the LAO-covered samples in Figs.~\ref{fig4} and \ref{fig3}. First of all, we can exclude that photoinduced transfer of the STO from the quantum-paraelectric to a bulk ferroelectric state causes the transient SHG increase. On the one hand, laser illumination is known to suppress rather than promote ferroelectricity \cite{Yamada2005}. On the other hand, the isotropy of the bulk would support multi-domain formation and, thus, cancelation of the net SHG signal. Instead, analysis of the STO dynamics in Fig.~\ref{fig2} reveals a simple model explaining our SHG data in a consistent way. Its mechanisms are pictorially outlined in Fig.~\ref{fig5}.

(i) As \textit{process 1} we assume that the pump beam creates an electron-hole-plasma that perturbs the interface polarity of the ground state. The plasma gives rise to two competing mechanisms, both well established in semiconductors. \textit{The first mechanism} is given by the drift of the photo-carriers in the presence of a preexisting equilibrium electric field, here the interfacial field $E^{pol}_z$, in order to screen it~\cite{Dekorsy1993,Chang2008}. We will refer to this as `screening drift' (Fig.~\ref{fig5}(b)). \textit{The second mechanism} originates from the different mobility of the photoexcited holes and electrons. The presence of the interface makes the diffusion anisotropic, thus inducing a net shift of the electron cloud with respect to the hole cloud. This gives rise to a local electric field at the interface (Fig.~\ref{fig5}(c)). In semiconductors this is known as `photo-Dember effect' \cite{Dember1931}. Henceforth, the parameters referring to electron-hole plasma generation and the two aforementioned \textit{charge-propagation}
  mechanisms will be labeled `cp'.

(ii) As \textit{process 2} we assume that the photogenerated carriers build up a transient polarization once they have reached the interface (Fig.~\ref{fig5}(d)). Mechanisms that may contribute to this polarization are interfacial charge trapping, polaron formation etc. They will be discussed in detail below. Henceforth, the parameters referring to the \textit{transient-polarization} mechanism will be labeled `tp'.

By combining `cp' and `tp' contributions and assuming an exponential behavior for their rise (`r') and decay (`d'), the temporal dynamic of $\Delta\bar{\chi}_{\rm loc/ext}$ is given
by the equation
\begin{eqnarray}\label{eq3}
 \Delta\bar{\chi}_{\rm loc/ext}&=&\Delta\bar{\chi}^{\rm cp}_{\rm loc/ext}
  (1-e^{-t/\tau^{\rm cp}_{\rm r}})
  e^{-t/\tau^{\rm cp}_{\rm d}}\nonumber\\
 &&+\Delta\bar{\chi}^{\rm tp}_{\rm loc/ext}
  (1-e^{-t/\tau^{\rm tp}_{\rm r}})
  e^{-t/\tau^{\rm tp}_{\rm d}}.
\end{eqnarray}
Here, $\tau^{\rm cp,tp}_{\rm r,d}$ and $\Delta\bar{\chi}_{\rm loc/ext}^{\rm cp,tp}$ denote the four time constants and the four SHG coupling coefficients, respectively.

Eq.~\ref{eq3} yields a total of eight parameters for describing the SHG data of each type of sample (LS2, LS6, STO). Only a subset of these parameters needs to be fitted, however, for describing the photocarrier dynamics. For example, $\tau^{\rm cp}_{\rm r}$ is resolution-limited and thus fixed at 0.1~ps. Furthermore, different SHG susceptibilities for the same type of process obey the same time constants [e.g.\ $\tau^{\rm tp}_{\rm d}(\Delta\bar{\chi}_{\rm ext})=\tau^{\rm tp}_{\rm d}(\Delta\bar{\chi}_{\rm loc}$)]. We thus obtain an excellent agreement between data and fit in Fig.~\ref{fig2}.

Henceforth, this fit procedure allows us to disentangle \textit{charge propagation} and \textit{transient polarization} based on their different dynamics. We therefore resume data analysis on the LAO-covered samples. We varied the photoinduced carrier density $N_{\rm ph}$, repeating the aforementioned fit procedure in each case. When $\tau$ did not display a clear dependence on $N_{\rm ph}$ we replaced it by a single value for all fits by a combined (global) fitting procedure. The resulting dependence of the SHG coupling coefficients and the time constants for the three types of samples are shown in Figs.~\ref{fig4} and \ref{fig3} in dependance of $N_{\rm ph}$. In the following, we will discuss how the competition of the aforementioned photoinduced processes can explain the rich dynamics displayed by Figs.~\ref{fig4} and \ref{fig3}.

The amplitudes of the \textit{charge-propagation} and \textit{transient-polarization} processes as defined in Eq.~\ref{eq3} are plotted in Fig.~\ref{fig4} as a function of photocarrier density $N_{\text{ph}}$. The solid lines are fits obtained using the following function of the photocarrier density $N_{\text{ph}}$: $\Delta \bar{\chi}/\bar{\chi}\propto N_{\text{ph}}/(1+N_{\text{ph}}/N^{sat}_{\text{ph}})$, where $N^{sat}_{\text{ph}}$ is a fit parameter describing saturation effects (see SM for their specific values). A superposition of two of these functions with opposite signs is needed for taking into account the zero-crossing behavior as a function of $N_{\text{ph}}$ for some amplitudes (LS2 in panel (a) and STO in panels (a) and (d)).

As mentioned, charge propagation is the result of screening drift, which enters the polarity balance with a negative sign ($E^{pol}_z$ is screened), and the photo-Dember effect, which enters the polarity balance with a positive sign for electrons more mobile than holes. In the STO and LS2 systems, the electronic reconstruction with charge injection and interface conductance does not occur. Hence, $E^{pol}_z$ is small or null and photo-Dember-like diffusion prevails over the screening drift --- net polarity increases. When the electronic interface reconstruction occurs, either inherently in the LS6 sample or as a photoinduced effect with increasing photocarrier density, $E^{pol}_z$ is large and the screening drift prevails --- net polarity decreases. The interplay of the two contributions explains the behavior yielded by the $d_{xz,yz}$-like subbands probed by $\bar{\chi}_{\rm ext}$ in Fig.~\ref{fig4}a. In contrast, the $d_{xy}$-like subband, probed by $\bar{\chi}_{\rm loc}$ in Fig.~\ref{fig4}c, shows only the decrease associated with the screening drift. Note that the photo-Dember effect describes an anisotropy in the charge propagation length that will only become obvious when an extended volume ($\sim\bar{\chi}_{\rm ext}$) is considered. In contrast, screening occurs all over the illuminated region and can therefore be perceived both when probing an extended region ($\sim\bar{\chi}_{\rm ext}$) and when probing a region atomically confined close to the interface ($\sim\bar{\chi}_{\rm loc}$).

As mentioned, the transient polarization builds up at the interface and increases the local electric field $E^{pol}_z$. Thus, the polarity change expressed by $\Delta\bar{\chi}^{\rm tp}(N_{\rm ph})$ is  positive in Figs.~\ref{fig4}b and \ref{fig4}d and saturates with the photocarrier density. As explained better below, this saturation behavior indicates the participation to the process of a finite number of pinning centers where photocarriers are trapped. A single exception to this understanding is the sign reversal of $\Delta\bar{\chi}^{\rm tp}_{\rm loc}$ that occurs with increasing photocarrier density in STO. We speculate that this might be related to the specific influence of Fr\"{o}hlich polarons which are known to form on the STO surface but to be suppressed at the LAO/STO interface~\cite{Cancellieri2016}.

Figure~\ref{fig3} scrutinizes the time scales of the \textit{charge-propagation} and \textit{transient-polarization} processes and reveals two main tendencies: (i) The lifetime of the  \textit{charge-propagation} contributions in the three types of samples is very different at low photocarrier density and quite similar at high photocarrier density. Thus, the difference between the insulating and the conducting samples reduces with increasing photocarrier density. A possible explanation for this would be that the conducting interface state that is inherently present in the LS6 sample is \textit{photoinduced} in the STO and LS2 samples. Then the behavior of the LS2 sample would approach that of the LS6 sample with increasing $N_{\rm ph}$, whereas the LS6 sample itself, where the conducting interface state is already present at $N_{\rm ph}=0$, would reveal minor $N_{\rm ph}$ dependence. This is exactly what we observe in Fig.~\ref{fig3}, and even though more evidence may be desirable, it strongly supports the scenario of a photoinduced conducting interface state.

(ii) The buildup time for the \textit{transient-polarization} contributions in all types of samples range from tens to hundreds of picoseconds and decreases by an order of magnitude when the photocarrier density is increased by a factor 400. This time is of the same order of magnitude of the time for the formation of a polaronic state~\cite{Yamada2013,Kohmoto2013} that hence becomes a good candidate for explaining the transition polarization state we observe. However, the lattice deformation accompanying the polaronic state must break the inversion symmetry to be visible in SHG. This might occur close to the interface where the photo-generated polarons may be trapped by defects and become polar for the presence of the interface. As already underlined, this trapping mechanism is proved by the saturation behavior of $\Delta \bar{\chi}^{\rm tp}$ as a function of $N_{\rm ph}$. It is further confirmed by the nonlinear behavior of $\tau^{\rm tp}_{\rm r}$ as a function of $N_{\rm ph}$. We fitted the dependence of $\tau^{\rm tp}_{\rm r}$ on $N_{\rm ph}$ with the relation
\begin{equation}\label{eq4}
 \tau^{\rm tp}_{\rm r}=\frac{\tau^{\rm tp}_{\rm r0}}{1+A N_{\rm ph}} + \tau_{\rm tr}.
\end{equation}
 where the time at null laser fluence, $\tau^{\rm tp}_{\rm r0}$, is probably fixed by the drifting time needed for the polarons to reach the interface, the term linear in $N_{\rm ph}$ describes the enhancement of the probability per unit time of filling a pinning center, and $\tau_{\rm tr}$ is related to the intrinsic polaron formation time and/or to the intrinsic trapping cross-section. The explicit values of these parameters are not needed for the present discussion and are reported in SM. The decrease of $\tau^{\rm tp}_{\rm r}$ with $N_{\rm ph}$ correlates well with the observed saturation behavior of $\Delta\bar{\chi}^{\rm tp}$ as a function of $N_{\rm ph}$ since a decrease of $\tau^{\rm tp}_{\rm r}$ accelerates filling of the pinning-centers.

The polaronic mechanism is further confirmed by the following argument. In conducting LAO/STO, the free-carrier mobility at room temperature is about 5 cm$^{\rm 2}{\rm V}^{\rm -1}{\rm s}^{\rm -1}$ \cite{Siemons2007}) which is in stark contrast to our measured drift time on the order of 100~ps. The drift mobility of polarons in STO, however, was found to be 0.04 cm$^{\rm 2}{\rm V}^{\rm -1}{\rm s}^{\rm -1}$, i.e., two orders of magnitude slower \cite{Keroack1984}. Assuming an electric field generated by the quantum well of about 40 mV$\cdot$nm$^{\rm -1}$~\cite{Lee2016} thus yields a polaron drift velocity of about 16$\cdot{10}^{\rm 10}{\rm nm}$ s$^{\rm -1}$. With a pump-laser penetration depth of 26~nm we find a transit time of the polarons toward the interface of about 160 ps, fully compatible with our measured values. We note that 160 ps is a lower limit since the electric field becomes weaker and weaker far from the interface where the polaron motion has mainly a diffusive character.

In conclusion, we have shown that photodoping can transiently change the polarity of the LAO/STO interface by more than 50\% within a few picoseconds. The polarity change is the consequence of competing charge-carrier-dynamical processes, namely \textit{charge propagation} in the form of screening drift and photo-Dember effect on the one hand and interfacial \textit{transient polarization} buildup on the other hand. The latter is probably due to the formation of polarons that drift toward the interface where they are eventually trapped.

We acknowledge funding from the European Union (FP7-PEOPLE-2012-CIG, Grant Agreement No. PCIG12-GA-2012-326499-FOXIDUET), the Deutsche Forschungsgemeinschaft (SFB 608 and TRR 80).

\bibliographystyle{plain}

\clearpage

\onecolumngrid
\appendix
\noindent\textbf{SUPPLEMENTARY INFORMATION}
\vspace{1 EM}

\section*{SHG theory for surfaces and interfaces}

In the following we provide a detailed derivation of the Eq. 1 of the main article. These details can be found here and there in our previous articles on static SHG of LAO/STO interfaces. We gather them here for the reader convenience. The SHG susceptibility
is given by~\cite{Deluca2014}:
\begin{equation}\label{eq1}
     \chi(t) = \int e^{2\omega}_{i} L^{{\rm out}}_{ii}
     \left[
           \chi_{ijk}^{(2)}(z)+\chi_{ijkz}^{(3)}(z)E^{pol}_z(t,z)
     \right]L^{{\rm in}}_{jj}L^{{\rm in}}_{kk}e^{\omega}_{j} e^{\omega}_{k} dz,
\end{equation}
where $z$ is the coordinate along the interface normal, $\mathbf{e}^{2\omega,\omega}$ are the electric-field unit vectors in vacuum. In Eq.~\ref{eq1}, $\chi_{ijk}^{(2)}$ describes the structural symmetry breaking occurring at any interface, and $\chi_{ijkl}^{(3)}$ with $l=z$ parameterizes the coupling to the electric field $E_z^{\rm pol}$ that is generated by the space-charge region at the LAO/STO or STO/air interface~\cite{Rubano2013a}. In the latter case a space-charge region may be created by charged surface defects. An optical excitation will drive the space-charge distribution out of equilibrium causing $E_z^{\rm pol}$ to become time-dependent. In general, $\chi_{ijk}^{(2)}$ and $\chi_{ijkl}^{(3)}$ might be also time-dependant because of the transient redistribution of the electronic populations by the pump beam, also known as state-filling effect~\cite{Sabbah2002}. However, as explained below, our measurements of photoinduced reflectivity indicate that these effects are negligible if compared to the free-carrier effects accounted by $E_z^{\rm pol}$, hence we will consider them as a constant.

$\mathbf{L}$ is a diagonal tensor giving the Fresnel transformation matrices that parameterize the propagation between the different media. The complete expression for the three non-zero components of this tensor can be found in Ref.~\onlinecite{Paparo2013}. These factors may vary in time because of changes in the refractive index. However, as explained below, we can assume that the Fresnel factors vary too little for explaining the absolute SHG signal variations or, in any case, they vary approximately in the same way for all the samples to account for the different dynamics observed among the samples. For all these reasons we neglect the Fresnel factors in our analysis.

In Eq.~\ref{eq1} the $z$ integral extends across the entire thickness of the polar layer where inversion symmetry is
broken by the presence of the interface~\cite{Rubano2011} or by the photoinduced charge redistribution. We underline here that,
unlike the standard optical techniques, the probing depth of SHG is not {\it a priori} fixed, but depends on the spatial extension
of the polar asymmetry at the specific interface under study. Since we measure the temporal, but not the spatial dependence of the LAO/STO
interface dynamics, we execute the $z$ integration in Eq.~(\ref{eq1}) and resort to space-averaged susceptibilities.
\begin{eqnarray}\label{eq2}
     \chi_{ijk}(t)=e^{2\omega}_{i}\left[\bar{\chi}_{ijk}+\Delta\bar{\chi}_{ijk}(t)\right]e^{\omega}_{j} e^{\omega}_{k},
\end{eqnarray}
where $\bar{\chi}_{ijk}=\int\chi^{(2)}_{ijk}dz+V_0\chi^{(3)}_{ijkz}$ and $\Delta\bar{\chi}_{ijk}(t)=\Delta V(t) \chi_{ijkz}^{(3)}$. Here we have introduced the ground state energy depth, $V_0$, of the quantum well induced by $E^{pol}_z(z,t<0)$ at the LAO/STO (STO/air) interface, and its light-induced variation, $\Delta V(t)$. By using suitable polarization combinations of the light all the elements of the $\overleftrightarrow{\chi}$ tensor can be measured.
\begin{figure}[h]
 \includegraphics[angle=0,scale=1]{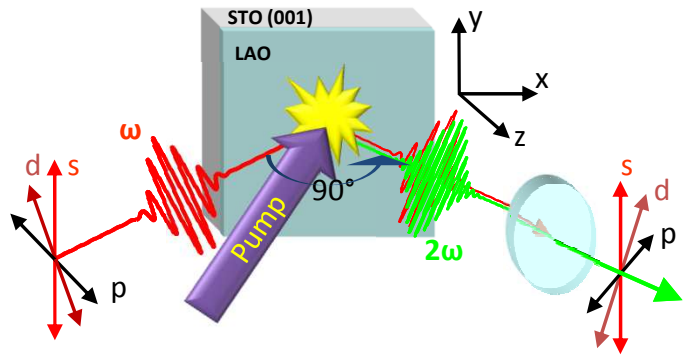}
 \caption{SHG pump-probe layout and geometry of the input/output light polarizations.}
\label{FigS1}
\end{figure}
As detailed in Ref.~\onlinecite{Paparo2013}, only three SHG contributions need to be distinguished for probing the LAO/STO interface. The susceptibility $\chi_{xxz}$ (and thus $\bar{\chi}_{xxz}$, $\Delta\bar{\chi}_{xxz}$) couples to the transition from the O$^{2-}(2p)$-dominated STO valence band to the lowest STO conduction sub-bands, mainly formed by the Ti$^{4+}(3d_{xy})$ orbital. In complement, $\chi_{zxx}$ (and thus $\bar{\chi}_{zxx}$, $\Delta\bar{\chi}_{zxx}$) represents the transition to the Ti$^{4+}(3d_{xz,yz})$ orbitals. The third component, $\chi_{zzz}$, is not considered since it cannot be measured directly and couples to the same transition as $\chi_{xxz}$ anyway. Note that the subbands with a $d_{xy}$ character are more localised at the interface~\cite{Popovic2008,Delugas2011} whereas those with a $d_{xz,yz}$ character are more extended so that $\chi_{xxz}$ and thus $\Delta\bar{\chi}_{xxz}$ (henceforth `$\chi_{\rm loc}$' and `$\Delta\bar{\chi}_{\rm loc}$') and $\chi_{zxx}$ and thus $\Delta\bar{\chi}_{zxx}$ (henceforth `$\chi_{\rm ext}$' and `$\Delta\bar{\chi}_{\rm ext}$') predominantly probe the local state at the interface and the extended environment around it, respectively.

Finally, in Eq.~1 of the main article,  $\Delta\bar{\chi}_{\rm loc/ext}$ is split in the two contributions $\Delta\bar{\chi}_{\rm loc/ext}^{\rm cp}$ and $\Delta\bar{\chi}_{\rm loc/ext}^{\rm tp}$, that are respectively associated to the {\it charge propagation} and {\it transient polarization} mechanisms. For each of these mechanisms we assume a rise-decay exponential behavior as given in Eq.~1 of the main article.

\section*{Sample fabrication}
LAO films of two and six unit cells were deposited by pulsed-laser deposition on TiO$_2$-terminated STO(001) substrates with unit-cell control of the film thickness by high-energy electron diffraction. The samples, henceforth refered to as LS2 and LS6, were grown at $\approx 800^{\circ}$C in 1 $\times$ 10$^{-4}$ mbar oxygen atmosphere and then cooled at this pressure to room temperature. LS2 and LS6 represent
the LAO/STO samples with an initial insulating and conducting interface, respectively, exhibiting sheet conductance at
300~K of $<10^{-9}$~$\Omega^{-1}$ (LS2) and \mbox{$\approx10^{-4}$~$\Omega^{-1}$} (LS6). Our pump beam excites photocarriers at 4.35~eV, i.e., above the direct band gap of SrTiO$_3$ of about 3.75~eV~\cite{Goldschmidt1987}, but not across the bulk band of LAO, so that most of the observed dynamics resides in STO. We therefore include the air/STO interface
dynamics of a TiO$_2$-terminated STO(001) substrate in our investigation.

\section*{SHG pump-probe experiment}
The pump-probe layout is shown in Fig.~1. As light source we used two phase-locked optical parametric amplifiers pumped by a single amplified Ti:Sa laser system (130~fs pulse duration, 1~kHz repetition rate, 1.55~eV nm central photon energy). The absorption length of the pump beam at a photon energy of 4.35~eV is 26~nm~\cite{Cohen1968}.
\begin{figure}[h]
 \includegraphics[angle=0,scale=0.65]{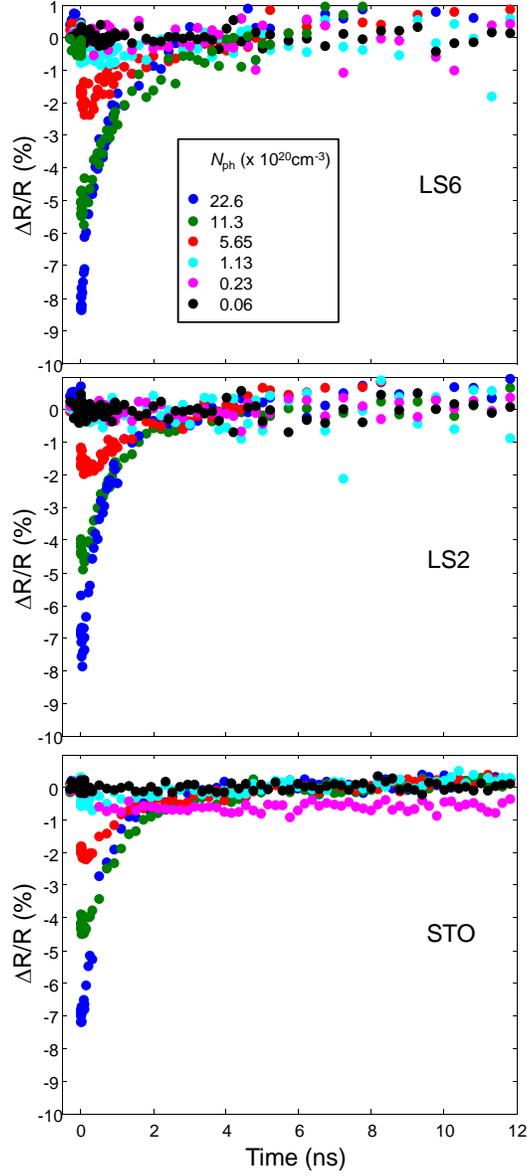}
 \caption{Pump-induced changes of reflectivity as a function of the photocarrier density for the samples: LS6 (top panel), LS2 (middle panel), and STO (bottom panel).}
\label{FigS2}
\end{figure}
It was focused onto an area of $280\pm 14$~$\mu$m. Because of the very small pump-pulse energy of 0.1 nJ-2 $\mu$J sample heating or deterioration are negligible. For probing the lowest cross-bandgap electronic states of the STO, the probe beam was tuned to an SHG photon energy of 3.75~eV. In order to subtract any long-term background the pump beam is chopped at 500 Hz and the difference of the SHG signal of two consecutive probe pulses is measured. This allows an accurate determination of the ultrafast response, while any dynamics exceeding 1 ms cannot be recorded.

In Fig.~1 the geometry of the input/output light polarizations is shown. According to the four-fold rotational symmetry (4$mm$) of the interface,
we measured the only three independent non-vanishing polarization combinations: $p$-in $p$-out ($pp$), $s$-in $p$-out ($sp$) and $d$-in $s$-out
($ds$)~\cite{Paparo2013}. These polarization combinations provide information on the three independent non-vanishing $\chi_{ijk}$-components:$\chi_{zzz}$, $\chi_{zxx}$ (`$\chi_{\rm ext}$' in the manuscript), and $\chi_{xxz}$ (`$\chi_{\rm loc}$' in the manuscript). The last two terms are obtained directly from the $sp$ and $ds$ signal, respectively.

\section*{Photoinduced reflectivity change}
Our time-resolved reflectivity measurements, as reported in Fig. 2, allow to neglect any variation of the SHG signal due to photoinduced changes of the refractive index compared to the free-carrier effects that are accounted by $E_z^{\rm pol}$. These photoinduced changes of the refractive index might vary the Fresnel factors. Somehow related to these changes are those possible for $\chi_{ijk}^{(2)}$ and $\chi_{ijkl}^{(3)}$, that might change because of the transient redistribution of the electronic populations by the pump beam, also known as state-filling effect.

 For what concerns the Fresnel factors the refractive-index changes observed in the reflectivity measurements are small either as absolute value or as relative
 variations among all the samples. Let us consider the SHG signal variations reported in Fig.~1 of the main manuscript. At this laser fluence the maximum
 reflectivity change is about 2\% for all samples. By using the refractive indexes of STO~\cite{STOindex} and the reflection Fresnel
 transformations~\cite{Fresnelreflection} this leads to a variation of the refractive index of about 1.1\%, approximately the same for both the fundamental and
 the second harmonic wavelengths. Now, according to Eq.~A5 of Ref.~\cite{Rubano2011}, the Fresnel factor L$_{zz}$ at 2$\omega$ and L$_{yy}$ at $\omega$ enter
 once and twice, respectively, in the $\bar{\chi}_{\rm ext}$ signal shown in Fig.~1 of the main manuscript. By inserting the variations of the refractive
 index in the corresponding Fresnel factors given by Eq.~A4 of Ref.~\cite{Rubano2011} we find that L$_{zz}$ and L$_{yy}$ vary of about 2\% and 0.9\%, respectively.
 This accounts for a total variation of 3.8\% that is approximately one order of magnitude less than the maximum variation observed in LS6 and STO, and a
 factor of six in the case of LS2. If we repeat this calculation for the Fig.~2 of the main manuscript we find that the total variation accounted by the Fresnel
 factor is even less than 1\%, whilst the maximum observed variation of the SHG signal is about 40\%. Furthermore, the comparison among all the samples at a fixed
 fluence shows that the differences between the reflectivity spectra are not larger than 1\%, at the best. Again, by using the same calculation as before,
 this leads to a relative variation among the samples due to the Fresnel factors of less than 1\%. This allows us to assume that the Fresnel factors vary too little for explaining the absolute SHG signal variations or, in any case, they vary approximately in the same way for all the samples to explain the different dynamics observed among the samples at a fixed laser fluence. For all these reasons we neglect the Fresnel factors in our analysis.

 Similarly, the comparison of the reflectivity spectra between conductive and insulating samples allows us to neglect any state-filling effect based on the following argument. The surface density of mobile charges at the LS6 interface is about 10$^{14}{\rm cm}^{-2}$~\cite{Savoia2009} (we note that this number is a lower limit since it does not include the injected charges that become localized). In order to compare this number with the photocarrier density induced by our pump we should know the space extension of the electronic gas at the interface. On the value of the latter there is not a unanimous consensus. However, we may safely assume that, at room temperature, it is confined in a region of the order of 1 nm with a more extended tail of localized charges~\cite{Cantoni2012}. The latter number gives a mobile carrier density of about 10$^{20}{\rm cm}^{-3}$-10$^{21}{\rm cm}^{-3}$, that is comparable to the photocarrier density induced in STO. This shows that the LS6 sample is already strongly populated, also in absence of the pump. However, the reflectivity changes that we measure in the STO and LS6 samples are approximately the same. This excludes possible state-filling effects at the laser fluences used in our experiments. For these reasons we assume that $\chi_{ijk}^{(2)}$ and $\chi_{ijkl}^{(3)}$ are constant in time.

\section*{Fitting results}
The data of $\Delta \bar{\chi}/\bar{\chi}$ as a function of $N_{\text{ph}}$ are fitted by using the following relation $N_{\text{ph}}/(1+N_{\text{ph}}/N^{sat}_{\text{ph}})$ or a combination of two of them with opposite signs when a zero-crossing behavior is present. Except for an overall amplitude the most significant fitting parameter is $N^{sat}_{\text{ph}}$. For completeness the values of the latter are reported in Table~\ref{table1}.
 \begin{table}[h]
\begin{tabular}{|l|c|c|c|c|c|c|c|c|}
  \hline
   \multicolumn{1}{|l|}{Sample} &  \multicolumn{4}{|c|}{Extended}  & \multicolumn{4}{|c|}{Localized} \\
   \cline{2-9}
  &  \multicolumn{2}{|c|}{`cp'-term}  & \multicolumn{2}{|c|}{`tp'-term} &  \multicolumn{2}{|c|}{`cp'-term}  & \multicolumn{2}{|c|}{`tp'-term}\\
   \cline{2-9}

   & $N^{sat}_{1\text{ph}}$ & $N^{sat}_{2\text{ph}}$ & $N^{sat}_{1\text{ph}}$ & $N^{sat}_{2\text{ph}}$& $N^{sat}_{1\text{ph}}$ & $N^{sat}_{2\text{ph}}$ & $N^{sat}_{1\text{ph}}$ & $N^{sat}_{2\text{ph}}$\\
   %& ($10^{20}$cm$^{-3}$)& ($10^{20}$cm$^{-3}$)& ($10^{20}$cm$^{-3}$)& ($10^{20}$cm$^{-3}$)& ($10^{20}$cm$^{-3}$)& ($10^{20}$cm$^{-3}$)& ($10^{20}$cm$^{-3}$)& ($10^{20}$cm$^{-3}$)\\
  \hline
  LS6& 0.38 & -    & 0.01 & - & 2.81 &-& 0.23 & - \\
  LS2& 0.73 & 0.76 & 0.10 & - & 1.89 &-& 0.13 & - \\
  STO& 1.94 & 2.03 & 0.27 & - & 0.05 &-& 5.26 & 5.31 \\
  \hline
\end{tabular}
 \caption{Values of $N^{sat}_{\text{ph}}$ obtained through the fitting procedure explained in the text. The subscripts 1 and 2 correspond to the two fitting functions with opposite signs needed for taking into account the zero-crossing behavior as a function of $N_{\text{ph}}$ for some amplitudes (LS2 in panel (a) and STO in panels (a) and (d) of Fig. 4 of the main article). The tabulated values are given in units of $10^{20}$ cm$^{-3}$.}
\label{table1}
\end{table}
The dependence of $\tau^{\rm tp}_{\rm r}$ on $N_{\rm ph}$ is fitted with the relation
\begin{equation}\label{eq4}
 \tau^{\rm tp}_{\rm r}=\frac{\tau^{\rm tp}_{\rm r0}}{1+A N_{\rm ph}} + \tau_{\rm tr}.
\end{equation}
 where the fitting parameters are in principle $A$, $\tau^{\rm tp}_{\rm r0}$, and $\tau_{\rm tr}$. Given the limited dynamic range of our measurements and the
\begin{table}[h]
\begin{tabular}{|l|c|c|}
  \hline

   \multicolumn{1}{|l|}{Sample} &  $A$ (10$^{-20}$ cm$^3$ ) & $\tau^{\rm tp}_{\rm r0} (ps) $ \\

  \hline
  LS6& 0.15 - 0.21 & 407 - 453 \\
  LS2& 0.14 - 0.20 & 383 - 430 \\
  STO& 0.08 - 0.12 & 280 - 327 \\
  \hline
\end{tabular}
 \caption{Values of A and  $\tau^{\rm tp}_{\rm r0}$ obtained through the fitting procedure explained in the text.}
\label{table2}
\end{table}
 large scattering of our data, the latter cannot be determined precisely. Therefore we keep it fixed in our fits. We get always good fits by varying $\tau_{\rm tr}$ in the range 50 - 0 ps. We note that this timescale is compatible with that reported in Ref.~\onlinecite{Yamada2013}. The corresponding intervals obtained for $A$ and $\tau^{\rm tp}_{\rm r0}$ are reported in Table~\ref{table2}. We note that $\tau^{\rm tp}_{\rm r0}$ increases by going from bare STO to LS6.

\bibliographystyle{plain}

\end{document}